# Decoding Taste Information in Human Brain: A Temporal and Spatial Reconstruction Data Augmentation Method Coupled with Taste EEG

Xiuxin Xia, Yuchao Yang, Yan Shi, Wenbo Zheng and Hong Men

***Abstract*—For humans, taste is essential for perceiving food's nutrient content or harmful components. The current sensory evaluation of taste mainly relies on artificial sensory evaluation and electronic tongue, but the former has strong subjectivity and poor repeatability, and the latter is not flexible enough. This work proposed a strategy for acquiring and recognizing taste electroencephalogram (EEG), aiming to decode people's objective perception of taste through taste EEG. Firstly, according to the proposed experimental paradigm, the taste EEG of subjects under different taste stimulation was collected. Secondly, to avoid insufficient training of the model due to the small number of taste EEG samples, a Temporal and Spatial Reconstruction Data Augmentation (TSRDA) method was proposed, which effectively augmented the taste EEG by reconstructing the taste EEG's important features in temporal and spatial dimensions. Thirdly, a multi-view channel attention module was introduced into a designed convolutional neural network to extract the important features of the augmented taste EEG. The proposed method has accuracy of 99.56%, F1-score of 99.48%, and kappa of 99.38%, proving the method's ability to distinguish the taste EEG evoked by different taste stimuli successfully. In summary, combining TSRDA with taste EEG technology provides an objective and effective method for sensory evaluation of food taste.**

## I. INTRODUCTION

AT present, foods with low cholesterol, low salt, low sugar, and low fat are favored by consumers. When such products are developed, evaluating their color, smell, taste, and tissue state is often necessary, which involves psychology, physiology, statistics, and other fields [1]. Among them, food taste is an important factor affecting food sensory evaluation. However, most current taste sensory evaluation relies on artificial sensory evaluation and machine perception, which has many shortcomings. In artificial sensory evaluation, when many people evaluate food taste, it is difficult for untrained sensory evaluators to evaluate the taste qualitatively.

Even professionally trained sensory evaluators are difficult to make a unified quantitative evaluation of complex and diverse tastes due to the limitation of words or language descriptions [2]. In the evaluation based on machine perception, such as electronic tongue, the analysis results are limited by the number and type of sensors, making it difficult to analyze comprehensively. Therefore, exploring new means for objective, accurate, and quantifiable sensory evaluation of food taste is necessary.

The brain's neural activity can be reflected by an electroencephalogram (EEG), which can effectively explain various sensory and psychological states [3, 4]. Brain-computer interface technology has been widely used in neuroscience research [5], disease treatment [6], multimedia [7], and other fields due to its advantages of low cost and easy portability. Taste EEG combines the substance's taste information with human sensory perception. Therefore, the sensory evaluators' objective feelings can be qualitatively and quantitatively analyzed by taste EEG, which has the potential in food sensory evaluation. In the current study, Crouzet et al. demonstrated that taste EEG induced by salt, sweet, sour, and bitter could be decoded by time-resolved multivariate pattern analysis, and then they used linear classifiers based on L2 regulated logistic regression to distinguish different tastes [8]. Hashida et al. used adaptive Gabor transform to extract features from taste EEG, revealing the significant difference between sweet, salty, and water [9]. Chandran et al. used a support vector machine classifier to classify salt, sweet, sour, and bitter after extracting the mean, median, and power spectral density features of taste EEG [10]. However, in traditional machine learning methods, the feature extraction and classification process are separated, which wastes time and makes it difficult to handle classification tasks flexibly. Even though the best classification features are obtained through continuous attempts, and the performance is partially improved, the features designed by humans may ignore the important information in the raw EEG.

In recent years, convolutional neural network (CNN) has been widely used in image recognition [11], machine perception [12, 13], and EEG analysis [14] because it avoids complex feature engineering. And it shows that it is superior to traditional machine learning methods without domain knowledge. Lawhern et al. proposed an EEGNet, which can learn various interpretable features in BCI tasks by temporal and spatial convolution [15]. Seal et al. proposed a CNN-based DeprNet for depression EEG recognition to learn the different response modes between the left and right hemispheres [16]. Zhao et al. proposed a multi-branch 3D

This work was supported by the Science and Technology Development Plan of Jilin Province YDZJ202101ZYTS135. *(Corresponding author: Hong Men).*

Xiuxin Xia, Yuchao Yang, Yan Shi, Wenbo Zheng and Hong Men are with the School of Automation Engineering, Northeast Electric Power University, Jilin, 132012, China and the Institute of Advanced Sensor Technology, Northeast Electric Power University, Jilin 132012, China (e-mails: 1202200014@neepu.edu.cn; 2202200631@neepu.edu.cn; shiyan@neepu.edu.cn; Wenbo_Zheng1996@163.com; menhong@neepu.edu.cn).

CNN, which can fully use the features of each EEG's dimension and effectively recognize motor imagery EEG [17]. However, most CNN consists of many convolutional and fully connected layers, which require many samples to train model parameters. Especially for taste EEG tasks, due to the complex experimental process, it is often difficult to obtain sufficient samples to train the model.

Data augmentation increases training data by generating new samples based on raw data, which is a feasible idea to improve CNN's performance and avoid over-fitting caused by insufficient training data. Therefore, data augmentation technology is necessary for taste EEG recognition. In the current research, Wang et al. solved the problem of insufficient emotion EEG by adding Gaussian noise to raw emotion EEG to increase the number of samples [18]. Zhang et al. reorganized epileptic EEG by multi-segment cutting and splicing to solve the unbalanced number of samples in epilepsy sets [19]. Pei et al. augmented the motor imagery EEG with a brain area recombination technique [20]. Schirrmeister et al. expanded the number of motor imagery EEG samples through the sliding time window segmentation technology [21]. Al-Saegh et al. proposed CutCat to augment motor imagery EEG by cutting specific periods from two samples and then connecting these cut parts [22]. Although the above methods have been proven feasible, they also have limitations. For example, adding Gaussian noise to raw EEG may introduce too much noise interference, which will drown the effective EEG signal. Cutting and recombining raw EEG in temporal or spatial dimensions may destroy EEG's temporal or spatial features. The samples generated by sliding time window segmentation are highly repetitive, and it is easy to cause redundant information. The CutCat cut raw EEG along the temporal dimension according to a fixed period, limiting the diversity of the augmented data and making it difficult to augment the local features effectively. Based on the above problems, it is necessary to explore a comprehensive, diverse, and effective data augmentation method to reconstruct the temporal and spatial features of taste EEG.

In image recognition, Yun et al. proposed a CutMix, which cut a patch from one image and replaced it with a patch in another image's corresponding position to produce a new image [23]. The new image had the class labels of the two images, and the proportion of class labels was proportional to the size of its corresponding patch. The CutMix can augment an image's features in any position, which improves the augmented data's diversity. In addition, when the augmented data with double labels is used for classifier optimization, it is more beneficial for the classifier to distinguish correct and wrong predictions, enhancing the classifier's generalization ability. Inspired by the CutMix, we designed a data augmentation method that was beneficial to reconstruct taste EEG data's temporal and spatial features based on the characteristics of taste EEG. And we introduced the idea of double-label joint optimization to our method to improve the method's robustness. Recently, attention mechanisms have been introduced into CNN models because they can mine the key features in convolution processes and improve classifiers' performance [24-27]. To fully mine the important features of augmented taste EEG and effectively avoid information redundancy caused by network deepening, it is feasible to introduce an attention mechanism into our CNN model.

This work established an experimental paradigm of taste EEG acquisition and proposed a taste EEG recognition strategy to distinguish the four tastes effectively. The main contributions are as follows.

(1) A taste EEG experimental paradigm was designed to evoke taste EEG using a self-developed taste EEG evoker and acquire taste EEG under different taste stimulation.

(2) A Temporal and Spatial Reconstruction Data Augmentation (TSRDA) method was proposed to effectively augment the taste EEG by reconstructing the taste EEG's important features in temporal and spatial dimensions.

(3) A multi-view channel attention module was introduced into a designed CNN to extract the important features of the augmented taste EEG. In a word, the whole set of proposed methods effectively distinguished the differences between the four tastes, which provided technical support for taste sensory evaluation.

## II. Materials and Methods

### *A. Experimental setting and data preprocessing*

#### 1) Subjects

Before the experiment, notices were posted in the community to solicit subjects for a fee, and 20 right-handed and non-smoking subjects (10 males and 10 females, aged 20 to 30) volunteered to participate. In the pre-experiment, all subjects could distinguish between sweet, sour, bitter, and salty tastes. They had no recent flu or fever symptoms and no neurological or psychiatric disorders history. They signed an informed consent form and were told to wash their hair, brush their teeth (unscented toothpaste) on the morning of the experiment, and not to eat (except water) for two hours before the experiment. The research was in line with the revised Declaration of Helsinki, and the protocol had been approved by the Northeast Electric Power University Scientific Research Ethics and Science and Technology Safety and Committee.

#### 2) Materials and Instrument

This work used the four tastes of sour, sweet, bitter, and salty often in food as experimental materials to analyze the taste EEG caused by these four basic tastes. Aqueous solutions for the four tastes were prepared with the details shown in Table I. They were prepared as follows: sour (0.075 g food grade citric acid dissolved in 100 ml distilled water, 0.039 M), sweet (15 g sucrose dissolved in 100 ml distilled water, 0.44 M), bitter (3 g bitter melon powder dissolved in 100 ml distilled water) and salty (3.8 g sodium chloride dissolved in 100 ml distilled water, 0.65 M), in addition, a bottle of 100 ml distilled water was prepared. According to previous studies [28], the prepared solution can be effectively perceived and does not cause discomfort or disgust to the subject, which was also verified in the pre-experiment.

TABLE I
INFORMATION OF FOUR KINDS OF EXPERIMENTAL SAMPLES

| No. | Taste | Composition | Manufacturer |
| --- | --- | --- | --- |
| 1 | Sour | 0.039 M citric acid + 100 ml distilled water | Weifang Yingxuan Industrial Co., Ltd. |
| 2 | Sweet | 0.44 M sucrose + 100 ml distilled water | Yunnan Dianpeng Sugar Co., Ltd. |
| 3 | Bitter | 3 g bitter gourd powder + 100 ml distilled water | Haozhou Deyongtang Biotechnology Co., Ltd. |
| 4 | Salty | 0.65 M sodium chloride + 100 ml distilled water | Xiaogan Guangyan Huayuan Salt Manufacturing Co., Ltd. |

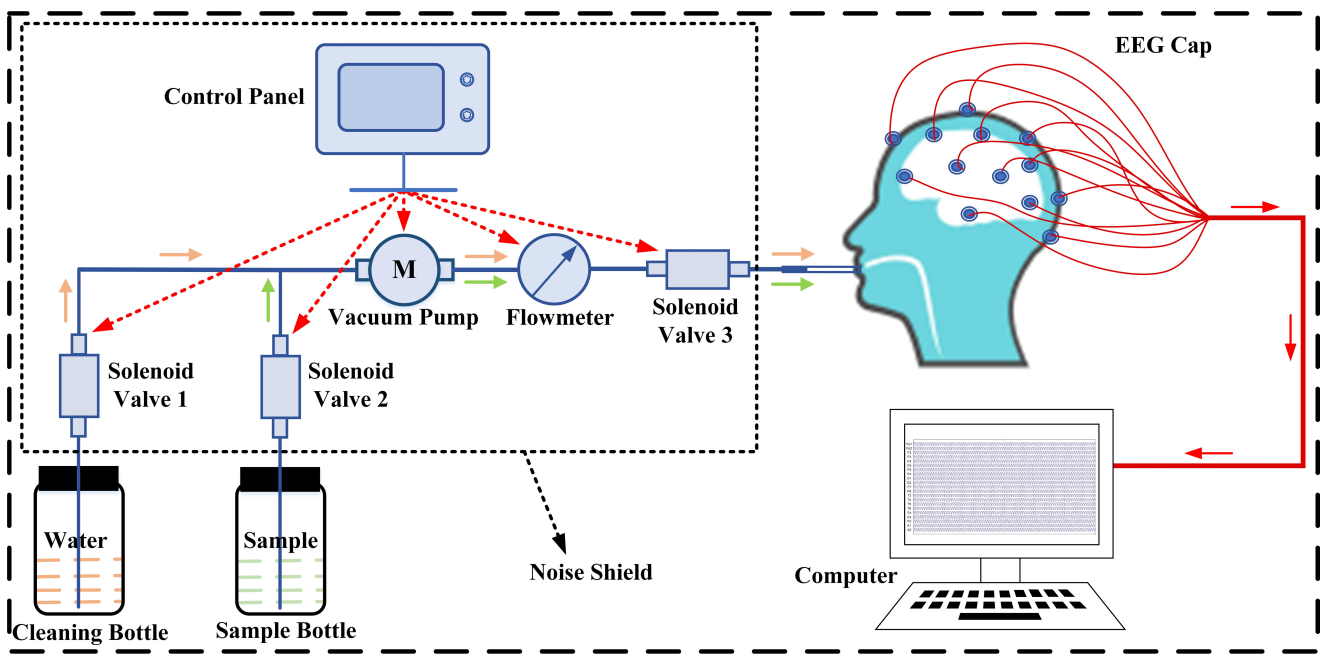

**Fig. 1.** The structure of the taste EEG evoker.

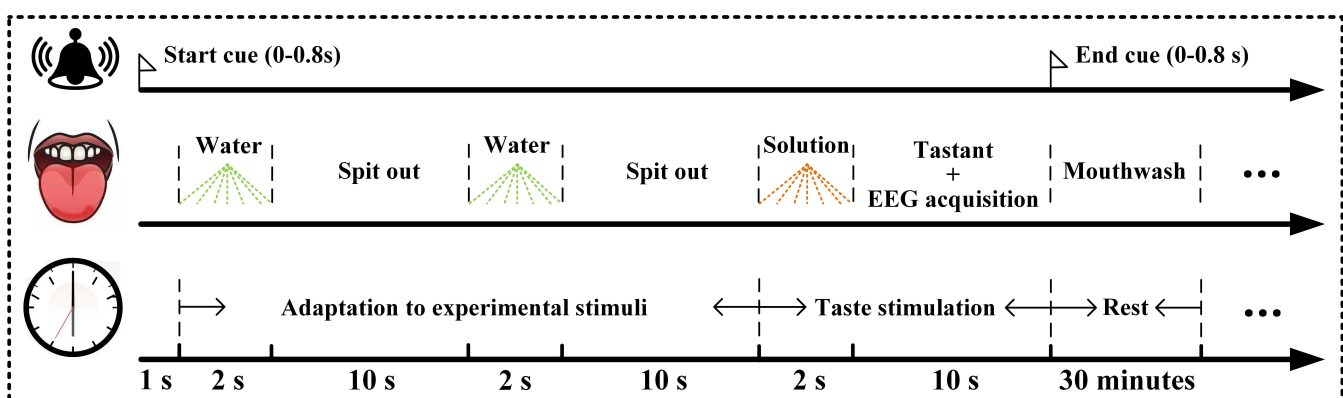

**Fig. 3.** The taste stimulation experiment procedure.

The self-developed low-noise, high-precision, and high-stability taste evoker was used to induce taste EEG. The structure of the taste evoker is shown in Fig. 1. It was mainly composed of an STM32F103ZE control panel (Dongguan Yehuo Electronic Technology Co., Ltd., China), S15S-53J micro vacuum pump (Chengdu Hailin Technology Co., Ltd., China), SFO-1037V-01 solenoid valve (Dongguan Sifan Electronic Technology Co., Ltd., China), AB32-S21P020C-11R liquid flowmeter (ODE Co., Ltd., (Hong Kong, China)) composition. Among them, the control panel was used to control the operation of the micro vacuum pump, the opening and closing of the solenoid valve, and the display of the liquid flow meter. The solenoid valve and vacuum pump were placed in a noise shield filled with soundproof cotton to eliminate noise further. In the experiment, the taste evoker had four working modes. Discharge of distilled water (Mode1): the control panel controlled solenoid valves 1 and 3 to open, and then the micro vacuum pump was activated to pump the distilled water to the nozzle quickly. Discharge of taste solution (Mode2): The control panel controlled solenoid valves 2 and 3 to open, and then the micro vacuum pump was activated to deliver the taste solution to the nozzle quickly. Distilled water stimulation (Mode3): Distilled water flowed gently to the subject's tongue at a rate of 0.25 ml/s for 2 s. Taste solution stimulation (Mode4): The taste solution flowed gently to the subject's tongue at a rate of 0.25 ml/s for 2 s.

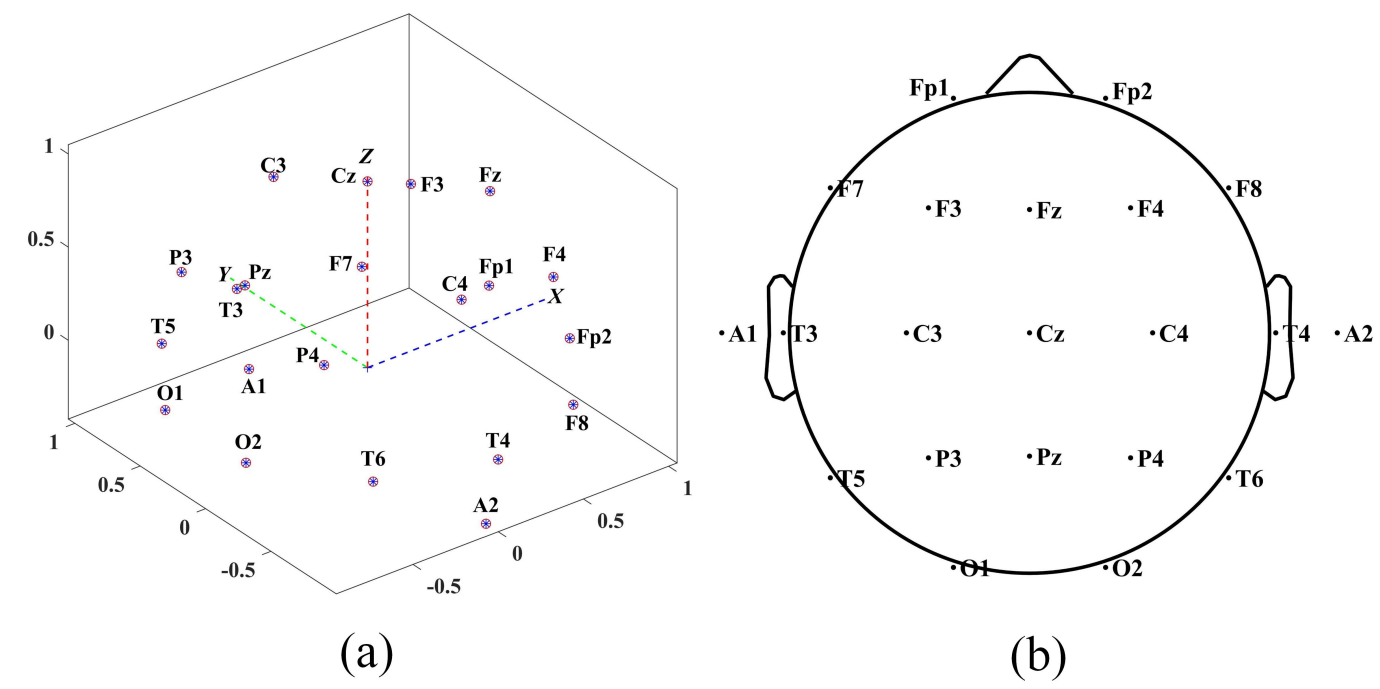

**Fig. 2.** The example positioning of 21-channel EEG electrodes: (a) 3D positioning of EEG electrodes, (b) 2D position projection.

Taste EEG was acquired by the NCERP-P EEG acquisition system (Shanghai NCC Electronics Co., Ltd., China) at a sampling frequency of 256 Hz. The 21 electrodes (Fz, Cz, Pz, T3, T4, C3, C4, Fp1, Fp2, F7, F8, T5, T6, O1, O2, F3, F4, P3, P4, A1, A2) of the EEG cap (GreentekPty. Ltd., China) were arranged according to the 10-20 system, and Fig. 2 shows the position distribution of 21-channel EEG electrodes.

**3) Experiment Process**

Before the experiment started, 100 ml of four sour, sweet, bitter, and salty solutions were prepared according to Table 1 and placed in 250 ml sampling bottles, respectively. The above four sampling bottles and the 250 ml cleaning bottle containing 100 ml of distilled water were placed in a heating dish, heated to 37.5 °C, and kept warm. To ensure no strong electromagnetic interference in the experimental environment. The environment temperature was controlled at 21 ± 2 °C through central air conditioning. For the experiment, each subject participated for four days, and the time of each day was 9 am to 11 am or 3 pm to 5 pm. Each subject will be randomly stimulated with four taste solutions during each day of the experiment. Before the taste stimulation started, the subject was put on the EEG cap, earplugs, and nasal plugs. And the chin rest was adjusted to support the subject's chin. The outlet tube of the taste evoker was placed 0.25-0.5 cm above the center of the subject's tongue to deliver the stimulation solution accurately. A taste solution was randomly selected, and the subject did not see the appearance of the solution. The two bottles containing the random taste solution and distilled water were connected to the taste evoker according to Fig. 1.

The process of the taste stimulation experiment is shown in Fig. 3. First, the taste evoker was set to Mode1 so that distilled water was quickly delivered to the nozzle. Then the taste evoker was set to Mode3, at which time the buzzer would emit a sound of appropriate volume that the subject could hear for 0.8 s as a prompt for the start of the experiment. After 1 s, 0.5 ml of distilled water will flow to the subject's tongue at a constant rate within 2 s, allowing the subject to adapt to the stimulation of the water flow. Then the subject spat out the liquid in the mouth and returned to the original position. After 10 s, the subject will receive the same stimulation again to strengthen the subject's adaptation to the water flow simulation. After 2 s, the subject spat out the liquid in the mouth and returned to the original position, during which the taste evoker was set to Mode2 to drain the tube of distilled water and quickly drain the taste solution to the mouth of the tube. After 10 s, the taste evoker will automatically switch to Mode4. At this time, 0.5 ml of taste solution will flow to the subject's tongue at a constant rate within 2 s, and then the subject will hold the liquid for 10 s to induce taste EEG, during which time the subject will avoid swallowing. After 10 s, the buzzer will sound to remind the end of the experiment. Then the subject rinsed their mouths thoroughly. After 30 minutes, proceed to the next taste stimulation experiment. The experiment of the day was not completed until the four kinds of taste stimulation were performed, and the other three days were parallel experiments with the same procedure.

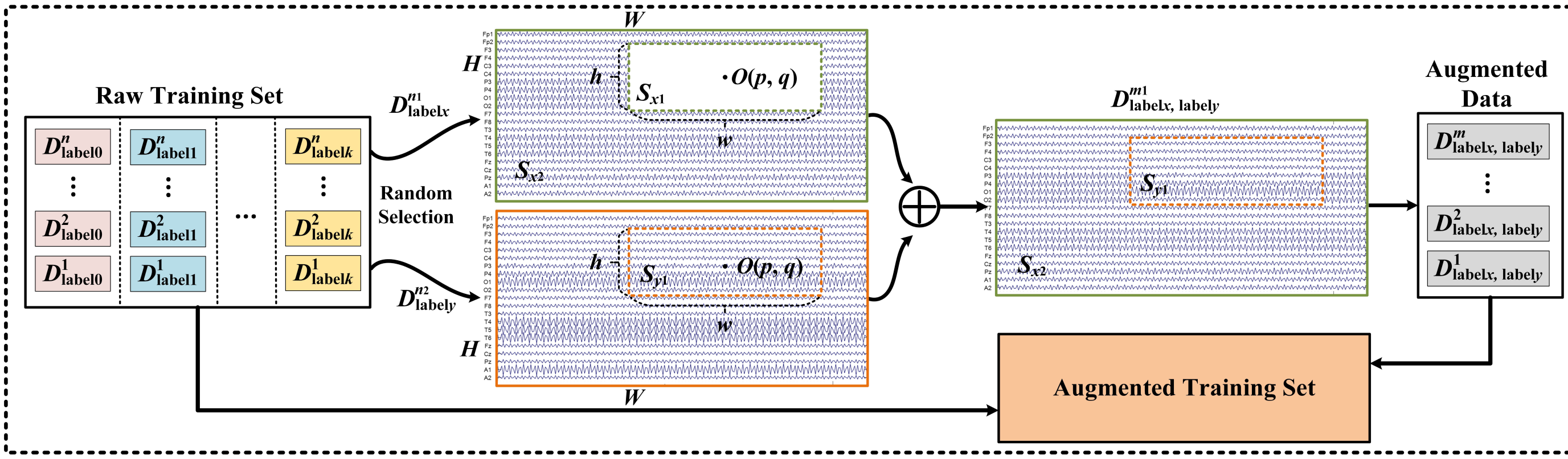

**Fig. 4.** The strategy of the TSRDA method.

**4) Taste EEG Preprocessing**

The taste EEG was preprocessed by Matlab (R2017b) and its built-in toolkit EEGLAB (version 2021). The processing flow is as follows.

(1) For each subject, four types of taste EEG were collected under four taste stimulation, each containing four parallel taste EEG data segments of 10 s. The size of the EEG samples was set to 2 s so that each data segment could produce five taste EEG samples labeled as 0, 1, 2, and 3 according to sour, sweet, bitter, and salty in that order. Thus, for 20 subjects, there were 1600 (20 × 4 × 4 × 5) EEG samples.

(2) After obtaining the samples, bandpass filtering between 0.5 and 50 Hz using a finite impulse response filter to remove low-frequency and high-frequency noise in the taste EEG. A notch filter removed the power frequency noise with the lower edge of the 49 Hz frequency passband and the upper edge of the 51 Hz frequency passband. Then, the sampling frequency of the samples was downsampled from 256 Hz to 128 Hz to reduce the sample size. Therefore, the size of an EEG sample was 256 × 21.

*B. Analytical Method*

**1) Temporal and Spatial Reconstruction Data Augmentation**

This work proposed the TSRDA for reconstructing taste EEG samples in the temporal and spatial dimensions. For the method, augmented training set samples with double labels were reconstructed by cutting and pasting raw training set samples. The ratio of the raw taste EEG samples' size corresponding to each label in the reconstructed samples was the ratio of the weights occupied by the labels. The augmented training set can provide sufficient training for the proposed CNN, improving the CNN's stability and generalization ability. The strategy of the TSRDA is shown in Fig. 4, and the main steps of its algorithm are as follows.

Step 1: Two samples $D_{labelx}^{n1}$ and $D_{labely}^{n2}$ are randomly selected from the raw training set, and the size of the samples is $W \times H$ (temporal dimension × spatial dimension).

Step 2: Cut out data blocks $S_{x1}$ and $S_{y1}$ of size $w \times h$ from the same locations in samples $D_{labelx}^{n1}$ and $D_{labely}^{n2}$, respectively, and $S_{x1}$ is covered by $S_{y1}$ to obtain the reconstructed data $D_{labelx,labely}^{m1}$.

Step 3: The reconstructed data has two labels, *labelx* and *labely*, whose weights are $1-\frac{\langle S_{y1}\rangle}{W\times H}$ and $\frac{\langle S_{y1}\rangle}{W\times H}$, respectively. Where $\langle\ \rangle$ represents the calculation of the size of the data block.

Step 4: When training the model, the two labels jointly optimize it according to their weights.

In this work, some detailed calculations in the TSRDA are described below. In step 1, $W$ = 256, $H$ = 21. In step 2, the upper left corner of the taste EEG was taken as (0,0) point coordinates, and the lower right corner coordinates were (256,21). The cutting of $S_{x1}$ and $S_{y1}$ needed to clarify its location and size. Among them, exploring the best location was particularly important for the TSRDA. The coordinate of the center of $S_{x1}$ and $S_{y1}$ data blocks were all $O(p,q)$, after parameter optimization, where $p=\lambda_p\times 256$, $q=\lambda_q\times 21$, $\lambda_p \sim Beta(1,1)$, $\lambda_q \sim Beta(2,1)$. The probability density function of the beta distribution *Beta* (*α, β*) is shown in Fig. 5. It is a uniform distribution when $\alpha = 1$ and $\beta = 1$. When $\alpha = 2$ and $\beta = 1$, the *x* closer to the value 1, the greater the probability density. When $\alpha = 1$ and $\beta = 2$, the *x* closer to the value 0, the greater the probability density. When $\alpha = 2$, $\beta = 2$, the *x* closer to the value 0.5, the greater the probability density. So when $\lambda_p \sim Beta(1,1)$ and $\lambda_q \sim Beta(2,1)$, the locations of the data blocks were evenly distributed in the temporal dimension and mainly distributed in the lower half of the EEG channel (O1-A2) in the spatial dimension. For the size, $w=\lambda_w\times 256$, $h=\lambda_h\times 21$, $\lambda_w \sim Beta(1,1)$, $\lambda_h \sim Beta(1,1)$. In most cases, the determined data block $(w\times h)$ would be included in the raw sample $(W\times H)$ according to the above cutting method. Still, sometimes some of the data blocks may be outside the sample. In this case, we only cut the part inside the sample for data reconstruction. In step 4, for the model's training, the reconstructed data $D^{m1}_{labelx,labely}$ was used for the forward propagation of the model to obtain the predicted value $output_{x,y}$, and then the cross-entropy loss values $loss_x$ and $loss_y$ were calculated using the labels *labelx* and *labely*, respectively. And then weighted and summed according to their respective weights to get the final loss value $loss_{x,y}$ for model optimization. Its calculation formula is as follows.

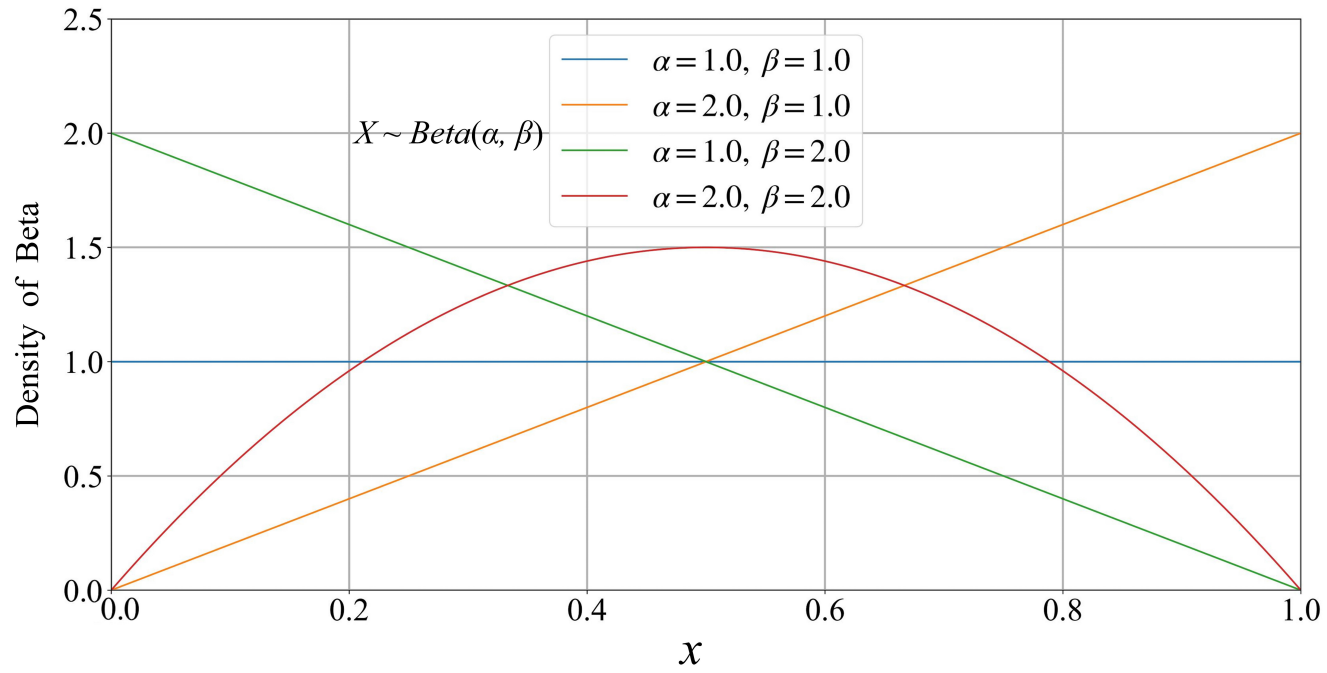

**Fig. 5.** The probability density function of the beta distribution.

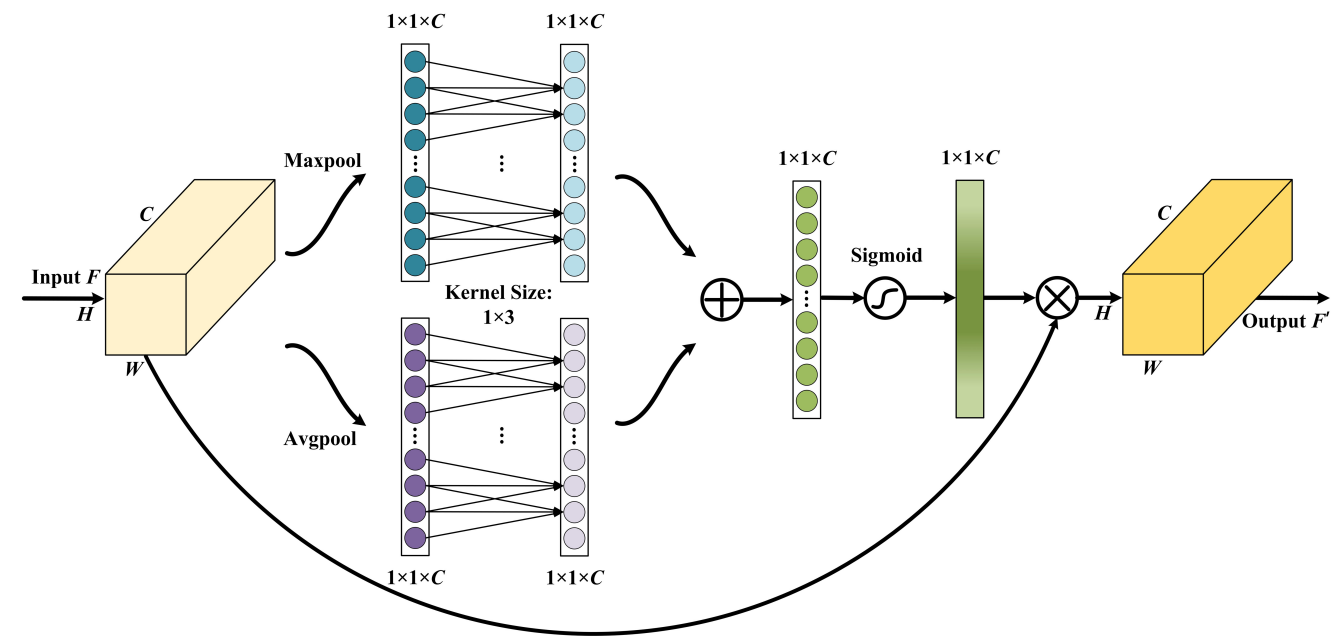

**Fig. 6.** The multi-view channel attention calculation process.

$$loss_x = \mathrm{model}(output_{x,y}, labelx) \tag{1}$$

$$loss_y = \mathrm{model}(output_{x,y}, labely) \tag{2}$$

$$loss_{x,y} = (1-\frac{\langle S_{y1}\rangle}{H\times W})\times loss_x + \frac{\langle S_{y1}\rangle}{H\times W}\times loss_y \tag{3}$$

### 2) Multi-view Channel Attention

CNN can extract the deep features of taste EEG, but increasing the network depth will also bring much redundant information. The channel attention mechanism can weaken the redundant information while focusing on key features. Therefore, we proposed multi-view channel attention, in which global average pooling and global max pooling were used to extract the taste EEG's overall features and main features, respectively. So the multi-view global feature information of taste EEG can be obtained. Then, one-dimensional convolution was performed on the multi-view feature information along the channel dimension, which effectively realized the cross-channel interaction of data.The calculation process of the multi-view channel attention is shown in Fig. 6, which mainly includes global pooling, one-dimensional convolution, information fusion, and nonlinear activation. Firstly, multi-view global feature information was obtained by global average pooling and global max pooling. Its calculation formula is as follows.

$$P_a = \frac{1}{H\times W}\sum_{i=1}^{H}\sum_{j=1}^{W}F_{ij} \tag{4}$$

$$P_m = \max(F_{ij}) \tag{5}$$

Where the input data $F\in\mathbb{R}^{H\times W\times C}$, $P_a$ and $P_m$ were the results of global average pooling and global max pooling of *F*, respectively.

Secondly, performed independent one-dimensional convolution operations on the $P_a$ and $P_m$, and realized cross-channel interaction of data through the sliding of convolution kernels. Its calculation formula is as follows.

$$R_a = \mathrm{conv1d}(P_a, C_{in}, C_{out}, kernel_size, padding) \tag{6}$$

$$R_m = \mathrm{conv1d}(P_m, C_{in}, C_{out}, kernel_size, padding) \tag{7}$$

Where the input channel $C_{in}$ was 1, the output channel $C_{out}$ was 1, the *kernel_size* was 3, and the *padding* was 1.

Thirdly, the data was added to realize information fusion, and then it was activated nonlinearly to obtain attention weight. Finally, the attention weight was multiplied by the original data *F* to get the multi-view channel attention result $F'$. Its calculation formula is as follows.

$$F' = F\times\delta(R_a + R_m) \tag{8}$$

where $\delta$ was a sigmoid function.

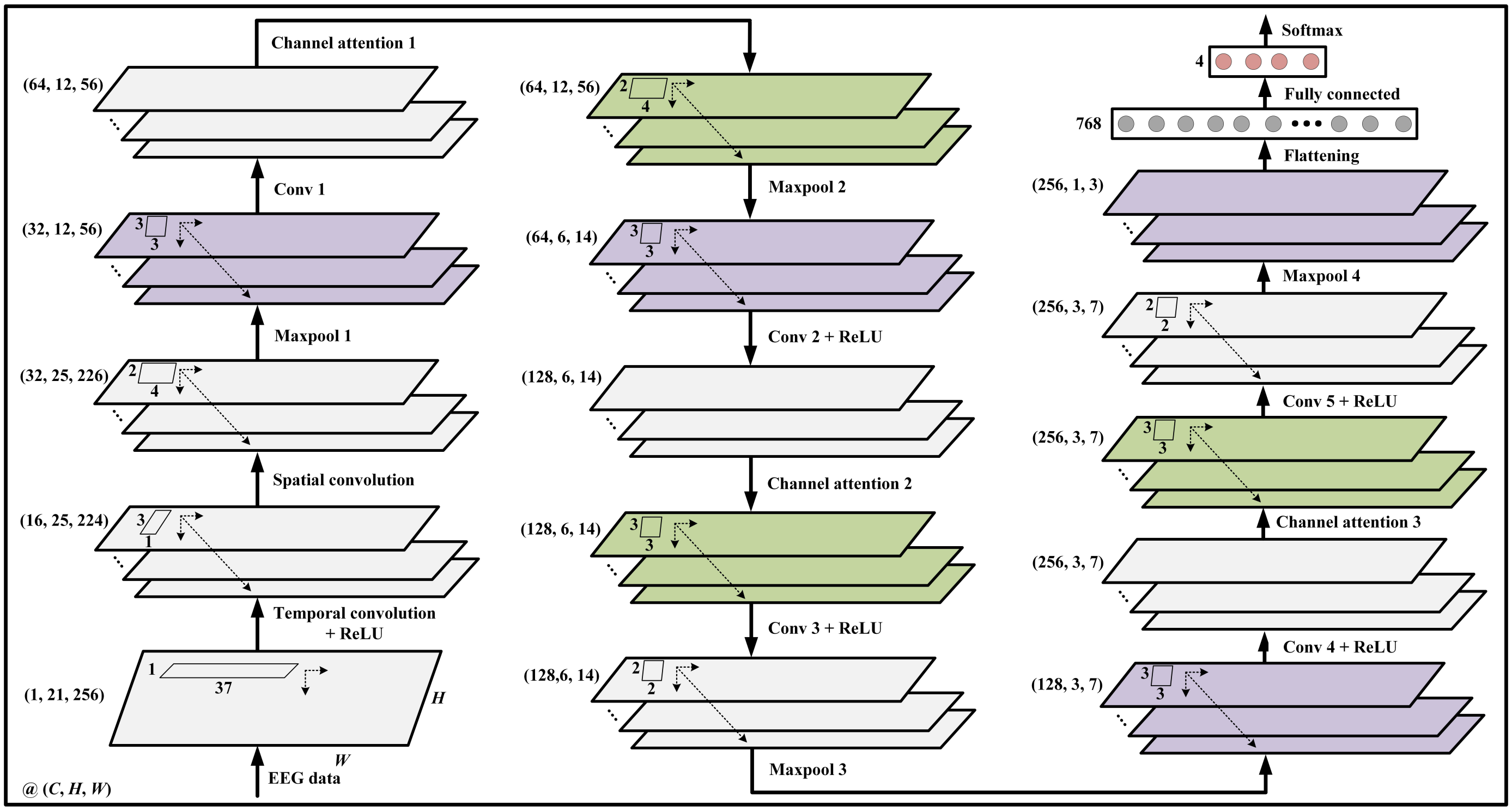

**Fig. 7.** The structure of the TSCNN-CA.

**3) TSCNN-CA Structural Design**

In recent years, CNN has been widely used in EEG classification tasks and has achieved superhuman performance. The CNN can effectively extract the high-dimensional features in EEG data through convolution and pooling. This work proposed a Temporal and Spatial Convolutional Neural Network with Channel Attention (TSCNN-CA) to extract the taste EEG's high-dimensional features and effectively recognize taste EEG. The structure of the TSCNN-CA is shown in Fig. 7. In the model, the input data size was 1 × 21 × 256. The temporal and spatial features of the taste EEG were mined through the temporal convolution and spatial convolution. In the temporal convolution, the kernel size was 1 × 37, the padding was 2, and the stride was 1. In the spatial convolution, the kernel size was 3 × 1, the padding was 1, and the stride was 1. The temporal domain contained more taste EEG information than the spatial domain, so the number of output channels was expanded 16 times and 2 times in the temporal and spatial convolution, respectively.

Then, a series of convolution and pooling was performed. In this process, higher-dimensional data features were extracted by convolution, which prompted the representation ability of the model. However, as data feature channels increased, redundant information was inevitably introduced. To avoid this problem and effectively mine important features in taste EEG data, multi-view channel attention (as shown in Fig. 6) was introduced after convolutions 1, 2, and 4. To improve the expressive ability of the network, ReLU was introduced as a nonlinear activation function after temporal convolution, convolutions 2, 3, 4, and 5. The details of the convolution and pooling operations are as follows. The kernel size in convolution 1 to convolution 5 was 3 × 3, the stride was 1, and the padding was 1. Only in convolution 1, convolution 2, and convolution 4, the number of data feature channels was doubled to extract the high-dimensional data features gradually. The maxpool was introduced to reduce the computational complexity and increase the receptive field of the model. The pooling kernel size of the maxpool 1 and 2 was 2 × 4, and the stride was 2 × 4. The pooling kernel size of the maxpool 3 and 4 was 2 × 2, and the stride was 2 × 2.

Finally, the feature map of size 256 × 1 × 3 was flattened and passed through the fully connected layer and softmax layer to obtain the prediction results.

## III. Results and Discussion

### *A. Hyperparameter Setting and Evaluation Methods*

In this work, 1600 taste EEG samples were randomly divided into the training set and test set according to 3 : 1, so the training set and the testing set contained 1200 and 400 taste EEG samples, respectively. Then, by the pre-experiment, the TSRDA was used to expand the training set by 3 times. Finally, the 4800 sample data of the augmented training set was used to train the TSCNN-CA, and the testing set was used to verify the performance of our proposed method. The augmented training set and testing set batch sizes were 64 and 32, respectively. In the model optimization process, the optimizer was Adam, the epoch was 100, the learning rate was 0.0005, and the weight decay was 0.001. To ensure the reliability of the experimental results, the average of the highest accuracy, the highest F1-score, and the highest kappa after each method's five independent experiments were used as the evaluation index.

TABLE II
COMPARISON OF OUR PROPOSED METHOD WITH STATE-OF-THE-ART METHOD

| Comparison Method | | | Accuracy (%) | F1-score (%) | Kappa (%) |
|---|---|---|---|---|---|
| Ref | Augmentation method | Classification model | | | |
| [18] | Add Gaussian noise (2019) | ResNet | 89.99 ± 1.75 | 89.68 ± 1.92 | 87.99 ± 2.45 |
| [19] | Time division and reorganization (2020) | Shallow CNN | 94.98 ± 3.19 | 94.69 ± 3.63 | 93.06 ± 4.42 |
| [20] | Brain area recombination (2021) | EEGNet | 95.30 ± 1.70 | 95.08 ± 1.94 | 93.60 ± 2.25 |
| [29] | Sliding time window segmentation (2019) | VGG-16 | 95.62 ± 2.30 | 95.31 ± 2.49 | 94.05 ± 3.11 |
| [21] | Sliding time window segmentation (2017) | Deep CNN | 97.52 ± 0.82 | 97.45 ± 0.98 | 96.66 ± 1.08 |
| [22] | CutCat (2021) | Deep CNN | 98.67 ± 0.52 | 98.57 ± 0.60 | 98.19 ± 0.71 |
| Ours | TSRDA | TSCNN-CA | **99.56 ± 0.21** | **99.48 ± 0.31** | **99.38 ± 0.29** |

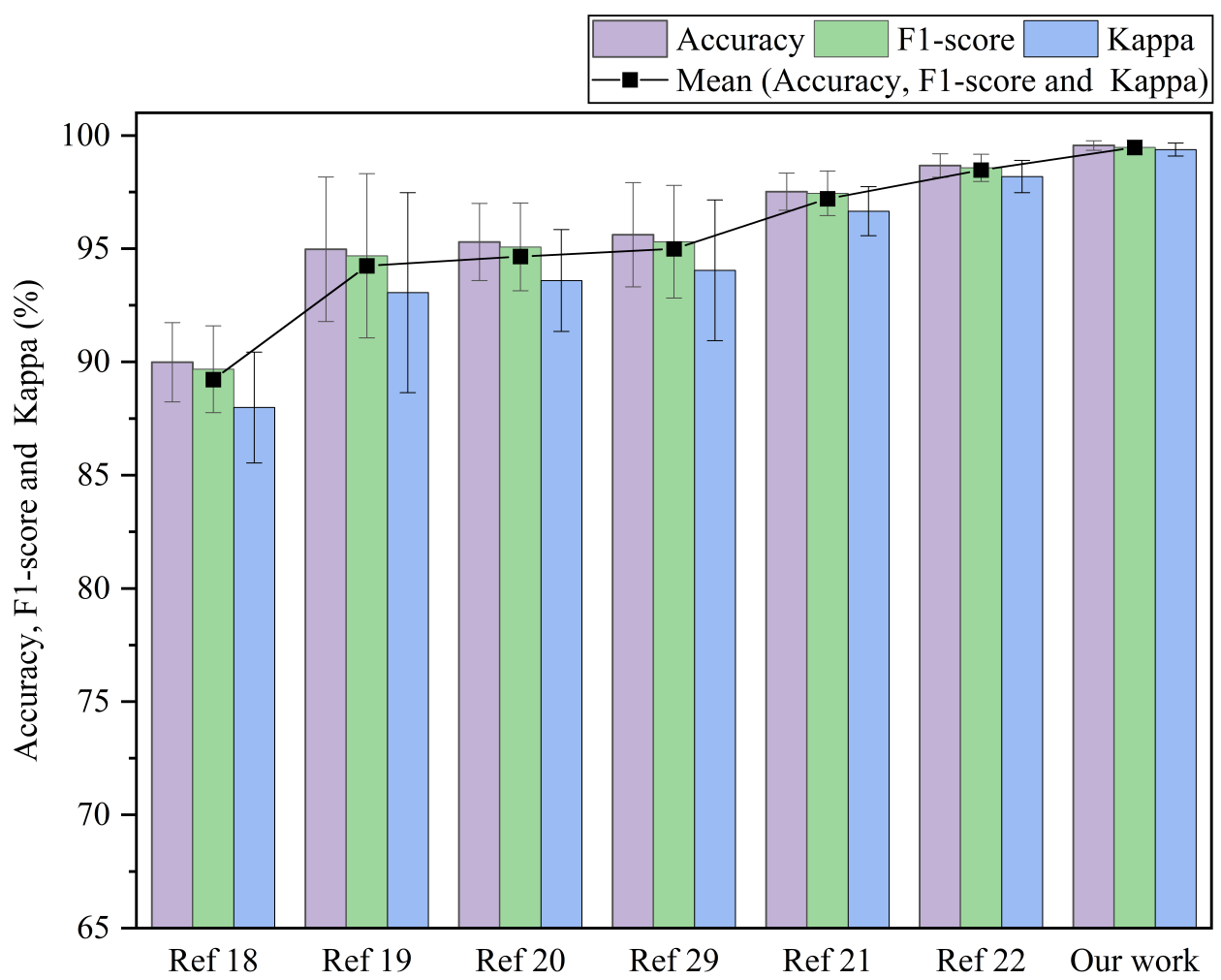

**Fig. 8.** Compared with the stability of the state-of-the-art method.

*B. Comparison with State-of-the-Art Models*

In our experiment, we compared our method with six state-of-the-art methods. In the comparison methods, the batch size, the optimizer, the epoch, the learning rate, and the weight decay of classification models were the same as the TSCNN-CA. The methods' brief descriptions are as follows.

Wang et al. augmented emotion EEG by adding Gaussian noise to raw emotion EEG and then used a ResNet model for classification [18]. Zhang et al. augmented epileptic EEG by cutting and splicing multiple EEG signals and then classified it with a Shallow CNN model [19]. Pei et al. proposed a brain area recombination method to augment motor imagery EEG, which improved the classification performance of an EEGNet [20]. Schirrmeister et al. augmented motor imagery EEG by sliding time window segmentation and classified it through a proposed Deep CNN model [21]. Al-Saegh et al. proposed a CutCat method to augment motor imagery EEG by cutting specific periods from two samples and then connecting these cut parts, finally realizing the classification of EEG signals by a Deep CNN model [22]. Xu et al. augmented motor imagery EEG by sliding time window segmentation and achieved a good classification of the motor imagery EEG through a proposed VGG-16 model [29].

Table II shows the comparison between the state-of-the-art method and our proposed method of combining the TSRDA with the TSCNN-CA. Generally speaking, our proposed method was superior to the other methods in taste EEG recognition, and its accuracy, F1-score, and kappa were 99.56%, 99.48%, and 99.38%, respectively. It can be seen that the method of augmenting the taste EEG by adding gaussian noise was the worst, which may be caused by introducing too much noise interference into the taste EEG. The methods such as time division and reorganization [19], brain area recombination [20], sliding time window segmentation [21, 29], and CutCat [22] have achieved good results. Still, their performance was limited because they can only augment the taste EEG from one dimension of temporal or spatial. Compared with the above methods, the accuracy of our proposed method was 9.57%, 4.58%, 4.26%, 2.04%, 0.89%, and 3.94% higher than the references [18], [19], [20], [21], [22] and [29] respectively, which fully demonstrated the advantages of our proposed TSRDA method to augment the taste EEG in both temporal and spatial dimensions. Meanwhile, the network depth of TSCNN-CA was similar to Deep CNN [21, 22]. It can be seen that TSCNN-CA not only ensured computational efficiency but also better mined the important features of taste EEG by multi-view channel attention.

The error bar chart shown in Fig. 8 reflects the accuracy, the F1-score, and the kappa of different methods. Our work was far superior to other methods in classification stability because it augmented and mined the important features of taste EEG while effectively avoiding the interference of redundant information and noise.

*C. Ablation Experiment*

**1) Data Augmentation Multiple**

For the TSRDA, too small a data augmentation multiple will lead to insufficient data augmentation, and too large will lead to information redundancy, affecting the accuracy and efficiency of taste EEG recognition. Therefore, it is necessary to determine the best data augmentation multiple.

In this work, the data augmentation multiple of the TSRDA was set to 1, 2, 3, 4, 5, and 6, respectively. The error bar chart shown in Fig. 9 reflects the performance of the TSCNN-CA under different data augmentation multiples of the TSRDA. On the whole, the performance of the TSCNN-CA under the different data augmentation multiple was different, but the overall performance was good. Compared with the raw training set, the TSRDA can improve the accuracy and stability of the TSCNN-CA. In addition, with the increase of data augmentation multiple from 1, 2 to 3, it can be seen that enough augmented data can fully train the classification model and improve the accuracy and stability. However, when the data augmentation multiple was greater than 3, the accuracy and stability decreased slightly due to the interference of redundant information. Therefore, setting the data augmentation multiple of the TSRDA to 3 ensured that the classification model was fully trained and avoided the problems of information redundancy and low recognition efficiency.

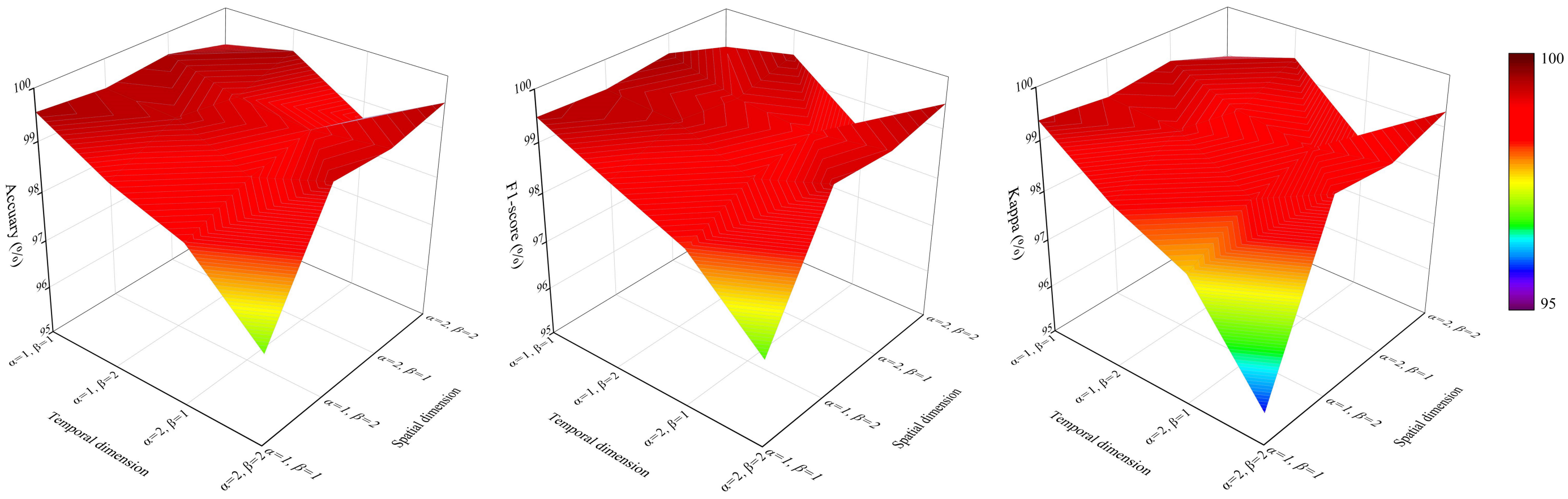

(a) (b) (c)

**Fig. 10.** The classification results of the TSCNN-CA under different reconstructed locations: (a) accuracy, (b) F1-score, (c) kappa.

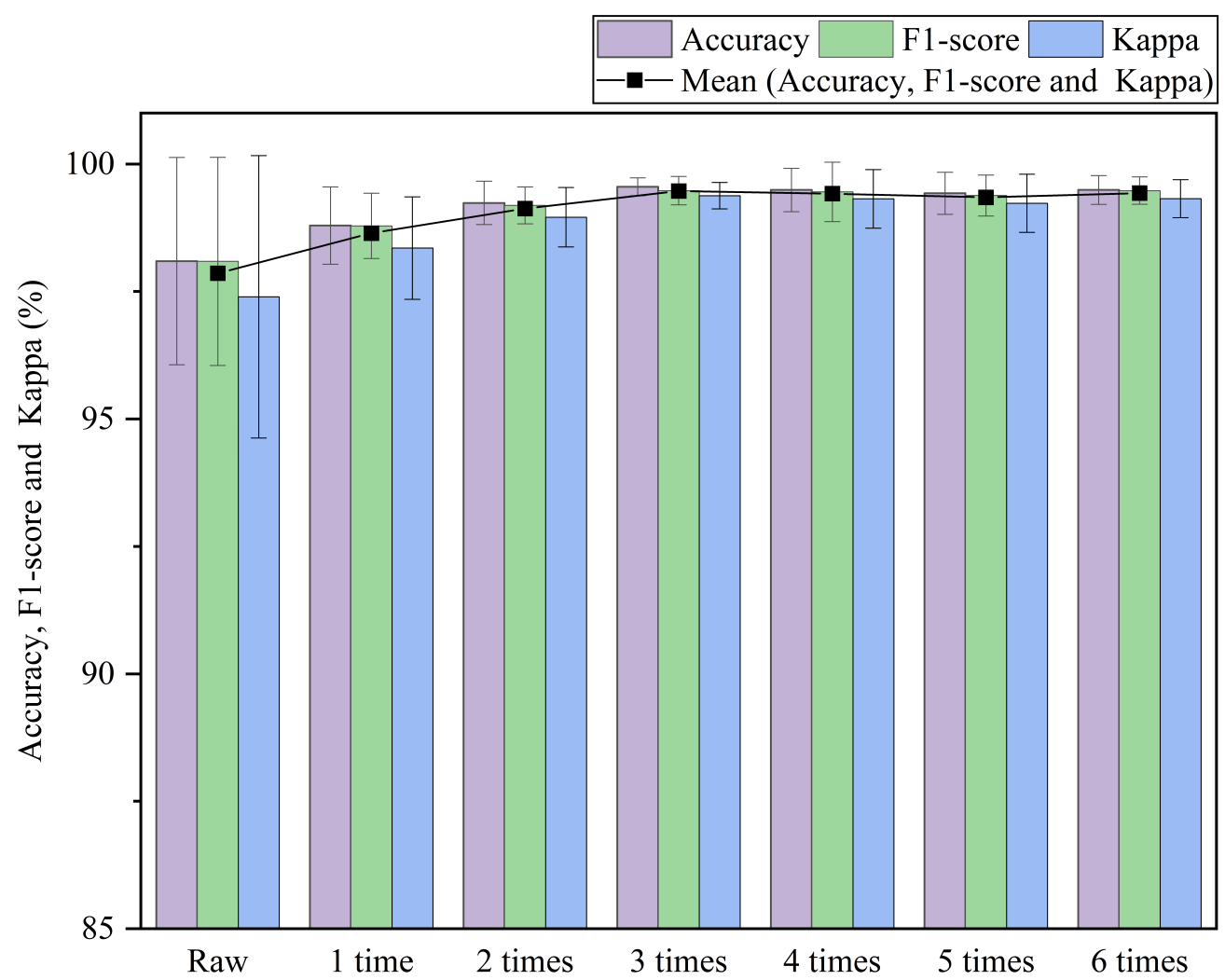

**Fig. 9.** The performance of the TSCNN-CA under different data augmentation multiples.

**2) Location of Temporal and Spatial Reconstruction**

For the taste EEG recognition tasks, subjects' differences and different taste stimulation will make the distribution of important features vary in temporal and spatial. Therefore, in the TSRDA, selecting the appropriate temporal and spatial reconstruction location can better reconstruct the important features of taste EEG, thus making it play a more significant role. We reconstructed the taste EEG (256 × 21) at different temporal and spatial locations. In this work, the $p$ and $q$ in the reconstructed data block's center coordinate $O(p, q)$ represented their locations in temporal and spatial, respectively. Where $p = \lambda_p \times 256$, $q = \lambda_q \times 21$. The probability density of the $\lambda_p$ was set as the distributions *Beta*(1, 1), *Beta*(1, 2), *Beta*(2, 1), and *Beta*(2, 2). According to Fig. 5, the four distributions represented that the reconstructed data blocks were distributed evenly across the temporal, mainly in the first half of the temporal, mainly in the second half of the temporal, and mainly in the middle of the temporal, respectively. Similarly, the probability density of the $\lambda_q$ was set as the distributions *Beta*(1, 1), *Beta*(1, 2), *Beta*(2, 1), and *Beta*(2, 2), which represented that the reconstructed data blocks were evenly distributed in the spatial, mainly in the upper half of the spatial, mainly in the lower half of the spatial, and mainly in the middle of the spatial, respectively. The TSRDA was used to augment the taste EEG data according to the reconstructed data blocks' different location distributions to explore the best temporal and spatial reconstruction location. The data augmentation multiple was set to 3, and the augmented data under reconstructed data blocks' different location distributions were used for the TSCNN-CA training.

The classification results of the TSCNN-CA under different reconstructed locations are shown in Fig. 10. In the temporal and spatial dimensions, when the probability density of the $\lambda_p$ and $\lambda_q$ respectively obeyed *Beta*(1, 1) and *Beta*(2, 1), the TSRDA achieved the best data augmentation effect for the taste EEG. In addition, the above results show that when the TSRDA was used to augment the taste EEG data, the uniform distribution of the reconstructed data blocks' locations in the temporal dimension was more conducive to fully augmenting the taste EEG's important features. It may indicate that the taste EEG's important features were distributed at each time of the temporal dimension. When the probability density of the $\lambda_p$ obeyed *Beta*(1, 1), the four distributions of the $\lambda_q$ in the spatial dimension achieved good results, probably due to the coupling between EEG channels. Notably, when the probability density of the $\lambda_q$ obeyed *Beta*(2, 1), the TSRDA had the best data augmentation effect. It may indicate that the taste EEG's important features were mainly distributed in the channels corresponding to the lower half of the spatial dimension.

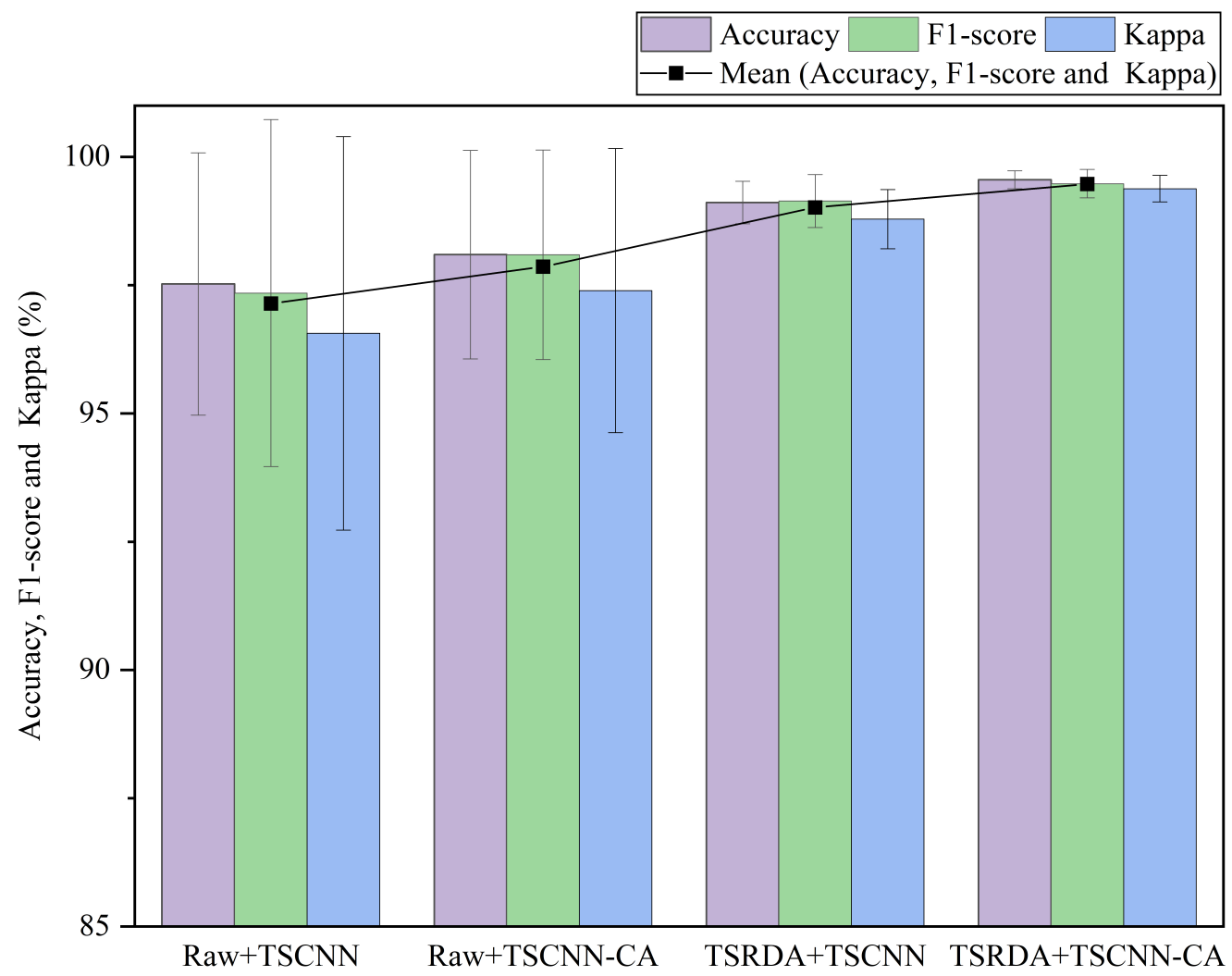

**Fig. 11.** The ablation experiment of the TSRDA and the multi-view channel attention.

**3) Data Augmentation and Channel Attention**

In this section, the proposed TSRDA and the multi-view channel attention in TSCNN-CA were ablated to analyze their effectiveness. The multi-view channel attention in TSCNN-CA was removed, which was a TSCNN. The comparison included Raw+TSCNN (without data augmentation and multi-view channel attention), Raw+TSCNN-CA (without data augmentation but with multi-view channel attention), TSRDA+TSCNN (with data augmentation but without multi-view channel attention), and TSRDA+TSCNN-CA (with data augmentation and channel attention). The ablation experiment result of the TSRDA and the multi-view channel attention is shown in Fig. 11. Comparing the classification results of the TSCNN and the TSCNN-CA before and after data augmentation, it can be seen that the proposed TSRDA effectively augmented the raw training set, improving the recognition accuracy and stability of the taste EEG. In addition, by comparing the Raw+TSCNN with Raw+TSCNN-CA and the TSRDA+TSCNN with TSRDA+TSCNN-CA, it can be seen that the introduction of multi-view channel attention can better mine the important features of the taste EEG, effectively avoiding the redundant information.

### *D. The Potential Applications*

In the taste EEG recognition task, due to the complexity of the experiment (time cost and money cost), it is difficult to obtain a large amount of the taste EEG, which increases the difficulty of taste EEG recognition. The proposed TSRDA can augment the taste EEG, which undoubtedly provides a new solution for the taste EEG recognition task. Although data augmentation is a good idea to solve the insufficient data, it sometimes will bring redundant information. The proposed multi-view channel attention method can effectively solve this problem. Furthermore, collecting EEG data in many other EEG tasks is often difficult. Our technical ideas of data augmentation and feature mining should be promising, and we will verify this in the future.

In addition, the proposed TSRDA determined the best reconstruction location by manual selection when reconstructing temporal and spatial features. Although it effectively augmented the taste EEG, it needed to be improved in convenience. In the future, we will introduce Grad-CAM [30] and other methods to find the most favorable feature location for reconstruction automatically. Undoubtedly, it will make our method more convenient and effective, broadening its application in EEG taste tasks. In short, we believe that the proposed experimental paradigm and the recognition strategy have great reference value for the research of taste EEG, providing technical support for the better application of taste EEG in food sensory evaluation.

## IV. Conclusion

In this work, the experimental paradigm of taste EEG acquisition was established. Then the proposed new EEG data augmentation method was combined with the designed attention convolutional neural network to realize taste recognition. The main conclusions were as follows.

(1) The established experimental paradigm of taste EEG acquisition has effectively collected the taste EEG induced by sour, sweet, bitter, and salty tastes.

(2) The proposed TSRDA method can effectively augment taste EEG by reconstructing its important features in temporal and spatial dimensions, which solves the problem of little taste EEG data.

(3) The proposed TSCNN-CA combined with the TSRDA has effectively recognized taste EEG. The TSRDA has ensured that TSCNN-CA is fully trained, and TSCNN-CA can effectively avoid the interference of redundant information in taste EEG recognition by the multi-view channel attention.

In summary, our proposed method is superior in data augmentation and classification model for taste EEG recognition compared with the state-of-the-art methods, which has effectively distinguished the differences between the four food tastes. We believe this method can be applie to the study of taste EEG, such as food sensory evaluation on taste.